\documentclass[prl,twocolumns,showpacs,groupedaddress,superscriptaddress,amsmath,amssymb,reprint, nofootinbib, nothanksinbib]{revtex4-2}

\usepackage{orcidlink}
\usepackage{hhline}
\usepackage{slashed}
\usepackage{graphicx}
\usepackage{graphics}
\usepackage{dcolumn}
\usepackage{bm}
\usepackage{amssymb}
\usepackage{enumerate}
\usepackage{lineno}
\usepackage{multirow} 
\usepackage{color}

\newcommand*{\Apr}{{A^\prime}}

\begin{document}

\title{Probing Light Dark Matter with positron beams at NA64}

\author{Yu.~M.~Andreev\orcidlink{0000-0002-7397-9665}}
\affiliation{Authors affiliated with an institute covered by a cooperation agreement with CERN}
\author{A.~Antonov\orcidlink{0000-0003-1238-5158}}
\affiliation{INFN, Sezione di Genova, 16147 Genova, Italia}
\author{D.~Banerjee\orcidlink{0000-0003-0531-1679}}
\affiliation{CERN, European Organization for Nuclear Research, CH-1211 Geneva, Switzerland}
\author{B.~Banto Oberhauser\orcidlink{0009-0006-4795-1008}}
\affiliation{ETH Z\"urich, Institute for Particle Physics and Astrophysics, CH-8093 Z\"urich, Switzerland}
\author{J.~Bernhard\orcidlink{0000-0001-9256-971X}}
\affiliation{CERN, European Organization for Nuclear Research, CH-1211 Geneva, Switzerland}
\author{P.~Bisio\orcidlink{/0009-0006-8677-7495}}
\thanks{Corresponding author}\email{pietro.bisio@ge.infn.it}
\affiliation{INFN, Sezione di Genova, 16147 Genova, Italia}
\affiliation{Universit\`a degli Studi di Genova, 16126 Genova, Italia}
\author{M.~Bond\'i\orcidlink{0000-0001-8297-9184}}
\affiliation{INFN, Sezione di Catania, 95125 Catania, Italia}
\author{A.~Celentano\orcidlink{0000-0002-7104-2983}}
\affiliation{INFN, Sezione di Genova, 16147 Genova, Italia}
\author{N.~Charitonidis\orcidlink{0000-0001-9506-1022}}
\affiliation{CERN, European Organization for Nuclear Research, CH-1211 Geneva, Switzerland}
\author{D.~Cooke}
\affiliation{UCL Departement of Physics and Astronomy, University College London, Gower St. London WC1E 6BT, United Kingdom}
\author{P.~Crivelli\orcidlink{0000-0001-5430-9394}}
\affiliation{ETH Z\"urich, Institute for Particle Physics and Astrophysics, CH-8093 Z\"urich, Switzerland}
\author{E.~Depero\orcidlink{0000-0003-2239-1746}}
\affiliation{ETH Z\"urich, Institute for Particle Physics and Astrophysics, CH-8093 Z\"urich, Switzerland}
\author{A.~V.~Dermenev\orcidlink{0000-0001-5619-376X}}
\affiliation{Authors affiliated with an institute covered by a cooperation agreement with CERN}
\author{S.~V.~Donskov\orcidlink{0000-0002-3988-7687}}
\affiliation{Authors affiliated with an institute covered by a cooperation agreement with CERN}
\author{R.~R.~Dusaev\orcidlink{0000-0002-6147-8038}}
\affiliation{Authors affiliated with an institute covered by a cooperation agreement with CERN}
\author{T.~Enik\orcidlink{0000-0002-2761-9730}}
\affiliation{Authors affiliated with an international laboratory covered by a cooperation agreement with CERN}
\author{V.~N.~Frolov}
\affiliation{Authors affiliated with an international laboratory covered by a cooperation agreement with CERN}
\author{A.~Gardikiotis\orcidlink{0000-0002-4435-2695}}
\affiliation{Physics Department, University of Patras, 265 04 Patras, Greece}
\author{S.~G.~Gerassimov\orcidlink{0000-0001-7780-8735}}
\affiliation{Authors affiliated with an institute covered by a cooperation agreement with CERN}
\affiliation{Technische Universit\"at M\"unchen, Physik Department, 85748 Garching, Germany}
\author{S.~N.~Gninenko\orcidlink{0000-0001-6495-7619}}
\affiliation{Authors affiliated with an institute covered by a cooperation agreement with CERN}
\author{M.~H\"osgen}
\affiliation{Universit\"at Bonn, Helmholtz-Institut f\"ur Strahlen-und Kernphysik, 53115 Bonn, Germany}
\author{M.~Jeckel}
\affiliation{CERN, European Organization for Nuclear Research, CH-1211 Geneva, Switzerland}
\author{V.~A.~Kachanov\orcidlink{0000-0002-3062-010X}}
\affiliation{Authors affiliated with an institute covered by a cooperation agreement with CERN}
\author{Y.~Kambar\orcidlink{0009-0000-9185-2353}}
\affiliation{Authors affiliated with an international laboratory covered by a cooperation agreement with CERN}
\author{A.~E.~Karneyeu\orcidlink{0000-0001-9983-1004}}
\affiliation{Authors affiliated with an institute covered by a cooperation agreement with CERN}
\author{G.~Kekelidze\orcidlink{0000-0002-5393-9199}}
\affiliation{Authors affiliated with an international laboratory covered by a cooperation agreement with CERN}
\author{B.~Ketzer\orcidlink{0000-0002-3493-3891}}
\affiliation{Universit\"at Bonn, Helmholtz-Institut f\"ur Strahlen-und Kernphysik, 53115 Bonn, Germany}
\author{D.~V.~Kirpichnikov\orcidlink{0000-0002-7177-077X}}
\affiliation{Authors affiliated with an institute covered by a cooperation agreement with CERN}
\author{M.~M.~Kirsanov\orcidlink{0000-0002-8879-6538}}
\affiliation{Authors affiliated with an institute covered by a cooperation agreement with CERN}
\author{V.~N.~Kolosov}
\affiliation{Authors affiliated with an institute covered by a cooperation agreement with CERN}
\author{I.~V.~Konorov\orcidlink{0000-0002-9013-5456}}
\affiliation{Technische Universit\"at M\"unchen, Physik Department, 85748 Garching, Germany}
\author{S.~V.~Gertsenberger\orcidlink{0009-0006-1640-9443}}
\affiliation{Authors affiliated with an international laboratory covered by a cooperation agreement with CERN}
\author{E.~A.~Kasianova}
\affiliation{Authors affiliated with an international laboratory covered by a cooperation agreement with CERN}
\author{V.~A.~Kramarenko\orcidlink{0000-0002-8625-5586}}
\affiliation{Authors affiliated with an institute covered by a cooperation agreement with CERN}
\affiliation{Authors affiliated with an international laboratory covered by a cooperation agreement with CERN}
\author{L.~V.~Kravchuk\orcidlink{0000-0001-8631-4200}}
\affiliation{Authors affiliated with an institute covered by a cooperation agreement with CERN}
\author{N.~V.~Krasnikov\orcidlink{0000-0002-8717-6492}}
\affiliation{Authors affiliated with an institute covered by a cooperation agreement with CERN}
\affiliation{Authors affiliated with an international laboratory covered by a cooperation agreement with CERN}
\author{S.~V.~Kuleshov\orcidlink{0000-0002-3065-326X}}
\affiliation{Center for Theoretical and Experimental Particle Physics, Facultad de Ciencias Exactas, Universidad Andres Bello, Fernandez Concha 700, Santiago, Chile}
\affiliation{Millennium Institute for Subatomic Physics at High-Energy Frontier (SAPHIR), Fernandez Concha 700, Santiago, Chile}
\author{V.~E.~Lyubovitskij\orcidlink{0000-0001-7467-572X}}
\affiliation{Authors affiliated with an institute covered by a cooperation agreement with CERN}
\affiliation{Millennium Institute for Subatomic Physics at High-Energy Frontier (SAPHIR), Fernandez Concha 700, Santiago, Chile}
\affiliation{Universidad T\'ecnica Federico Santa Mar\'ia and CCTVal, 2390123 Valpara\'iso, Chile}
\author{V.~Lysan\orcidlink{0009-0004-1795-1651}}
\affiliation{Authors affiliated with an international laboratory covered by a cooperation agreement with CERN}
\author{A.~Marini\orcidlink{0000-0002-6778-2161}}
\affiliation{INFN, Sezione di Genova, 16147 Genova, Italia}
\author{L.~Marsicano\orcidlink{0000-0002-8931-7498}}
\affiliation{INFN, Sezione di Genova, 16147 Genova, Italia}
\author{V.~A.~Matveev\orcidlink{0000-0002-2745-5908}}
\affiliation{Authors affiliated with an international laboratory covered by a cooperation agreement with CERN}
\author{R.~Mena~Fredes}
\affiliation{Millennium Institute for Subatomic Physics at High-Energy Frontier (SAPHIR), Fernandez Concha 700, Santiago, Chile}
\author{R.~Mena~Yanssen}
\affiliation{Millennium Institute for Subatomic Physics at High-Energy Frontier (SAPHIR), Fernandez Concha 700, Santiago, Chile}
\affiliation{Universidad T\'ecnica Federico Santa Mar\'ia and CCTVal, 2390123 Valpara\'iso, Chile}
\author{Yu.~V.~Mikhailov}
\affiliation{Authors affiliated with an institute covered by a cooperation agreement with CERN}
\author{L.~Molina Bueno\orcidlink{0000-0001-9720-9764}}
\affiliation{Instituto de Fisica Corpuscular (CSIC/UV), Carrer del Catedratic Jose Beltran Martinez, 2, 46980 Paterna, Valencia, Spain}
\author{M.~Mongillo\orcidlink{0009-0000-7331-4076}}
\affiliation{ETH Z\"urich, Institute for Particle Physics and Astrophysics, CH-8093 Z\"urich, Switzerland}
\author{D.~V.~Peshekhonov\orcidlink{0009-0008-9018-5884}}
\affiliation{Authors affiliated with an international laboratory covered by a cooperation agreement with CERN}
\author{V.~A.~Polyakov\orcidlink{0000-0001-5989-0990}}
\affiliation{Authors affiliated with an institute covered by a cooperation agreement with CERN}
\author{B.~Radics\orcidlink{0000-0002-8978-1725}}
\affiliation{Department of Physics and Astronomy, York University, Toronto, ON, Canada}
\author{K.~Salamatin\orcidlink{0000-0001-6287-8685}}
\affiliation{Authors affiliated with an international laboratory covered by a cooperation agreement with CERN}
\author{V.~D.~Samoylenko}
\affiliation{Authors affiliated with an institute covered by a cooperation agreement with CERN}
\author{H.~Sieber\orcidlink{0000-0003-1476-4258}}
\affiliation{ETH Z\"urich, Institute for Particle Physics and Astrophysics, CH-8093 Z\"urich, Switzerland}
\author{D.~Shchukin\orcidlink{0009-0007-5508-3615}}
\affiliation{Authors affiliated with an institute covered by a cooperation agreement with CERN}
\author{O.~Soto}
\affiliation{Millennium Institute for Subatomic Physics at High-Energy Frontier (SAPHIR), Fernandez Concha 700, Santiago, Chile}
\affiliation{Departamento de Fisica, Facultad de Ciencias, Universidad de La Serena, Avenida Cisternas 1200, La Serena, Chile}
\author{V.~O.~Tikhomirov\orcidlink{0000-0002-9634-0581}}
\affiliation{Authors affiliated with an institute covered by a cooperation agreement with CERN}
\author{I.~Tlisova\orcidlink{0000-0003-1552-2015}}
\affiliation{Authors affiliated with an institute covered by a cooperation agreement with CERN}
\author{A.~N.~Toropin\orcidlink{0000-0002-2106-4041}}
\affiliation{Authors affiliated with an institute covered by a cooperation agreement with CERN}
\author{A.~Yu.~Trifonov}
\affiliation{Authors affiliated with an institute covered by a cooperation agreement with CERN}
\author{M.~Tuzi\orcidlink{0009-0000-6276-1401}}
\affiliation{Instituto de Fisica Corpuscular (CSIC/UV), Carrer del Catedratic Jose Beltran Martinez, 2, 46980 Paterna, Valencia, Spain}
\author{P.~Ulloa\orcidlink{0000-0002-0789-7581}}
\affiliation{Center for Theoretical and Experimental Particle Physics, Facultad de Ciencias Exactas, Universidad Andres Bello, Fernandez Concha 700, Santiago, Chile}
\author{P.~V.~Volkov\orcidlink{0000-0002-7668-3691}}
\affiliation{Authors affiliated with an international laboratory covered by a cooperation agreement with CERN}
\author{V.~Yu.~Volkov\orcidlink{0009-0005-3500-5121}}
\affiliation{Authors affiliated with an institute covered by a cooperation agreement with CERN}
\author{I.~V.~Voronchikhin\orcidlink{0000-0003-3037-636X}}
\affiliation{Authors affiliated with an institute covered by a cooperation agreement with CERN}
\author{J.~Zamora-Sa\'a\orcidlink{0000-0002-5030-7516}}
\affiliation{Center for Theoretical and Experimental Particle Physics, Facultad de Ciencias Exactas, Universidad Andres Bello, Fernandez Concha 700, Santiago, Chile}
\affiliation{Millennium Institute for Subatomic Physics at High-Energy Frontier (SAPHIR), Fernandez Concha 700, Santiago, Chile}
\author{A.~S.~Zhevlakov\orcidlink{0000-0002-7775-5917}}
\affiliation{Authors affiliated with an international laboratory covered by a cooperation agreement with CERN}


\date{\today}

\begin{abstract}
We present the results of a missing-energy search for Light Dark Matter which has a new interaction with ordinary matter transmitted by a vector boson, called dark photon $\Apr$. For the first time, this search is performed with a  positron beam by using the significantly enhanced production of  $\Apr$ in the resonant  annihilation of positrons with atomic electrons of the target nuclei, followed by the invisible decay of $\Apr$ into dark matter. No events were found in the signal region with $(10.1 \pm 0.1)~\times~10^{9}$ positrons on target with 100 GeV energy. This allowed us to set new exclusion limits that, relative to the collected statistics, prove the power of this experimental technique. This measurement is a crucial first step toward a future exploration program with positron beams, whose estimated sensitivity is here presented.
\end{abstract}

\maketitle
Numerous cosmological and astrophysical observations unequivocally indicate that 85$\%$ of the matter of our Universe is made by a new form of matter, called ``Dark Matter'' (DM), gravitationally interacting with the ordinary matter described by the Standard Model (SM), but not directly emitting or absorbing light~\cite{Bertone:2016nfn,Liddle:1998ew,Arcadi:2017kky}. At present, the particle content of DM is unknown. Among the different theories that have been postulated to explain the DM microscopic properties, the Light Dark Matter (LDM) hypothesis assumes that DM is made by sub-GeV particles, interacting with the SM through a new force. LDM particles (here denoted as $\chi$) can be the lightest stable states of a new ``Dark Sector'' (DS), with its own particles and fields
~\cite{Arkani-Hamed:2008hhe,Strassler:2006im,Andreas:2011in}. 
A representative LDM model 
involves the existence of a new $U(1)_D$ hidden symmetry in Nature, associated to a massive gauge boson, also called ``dark photon'' ($\Apr$). The dark photon can kinetically mix with the ordinary photon, thus acting as a ``portal mediator'' between the DS and the SM~\cite{Holdom:1985ag,Knapen:2017xzo}. In this framework, the new Lagrangian term extending the SM, omitting the LDM mass term, reads:
\begin{equation}\label{eq:lagrangian}
 \mathcal{L} \supset -\frac{1}{4} F'_{\mu\nu}F'^{\mu\nu}
 +\frac{1}{2}m^2_\Apr \Apr_\mu \Apr^\mu
 -\frac{\varepsilon}{2}F_{\mu\nu}F'^{\mu\nu} -g_D \Apr_\mu J^\mu_D 
\end{equation}
where $m_\Apr$ is the dark photon mass, $F'_{\mu\nu}\equiv \partial_\mu A'_\nu - \partial_\nu A'_\mu$ is the dark photon field strength tensor, $F_{\mu\nu}$ is the SM electromagnetic field strength, $g_D\equiv\sqrt{4 \pi \alpha_D}$ is the dark gauge coupling, $J^\mu_D$ is the LDM current under $U(1)_D$, and $\varepsilon$ parametrizes the mixing strength. While it is reasonable to assume that at tree level $g_D \sim 1$, the range $\sim 10^{-4} - 10^{-2}$ ($\sim 10^{-6} - 10^{-3}$) is predicted for $\varepsilon$, if the kinetic mixing is generated at the one (two)-loop level~\cite{Essig:2010ye,delAguila:1988jz,ArkaniHamed:2008qp}. Cosmological arguments connected to the DM thermal origin in the early Universe provide a relation between the measured DM relic density and the model parameters. 
Specifically, by introducing the dimensionless parameter $y \equiv \alpha_D \varepsilon^2 \left(\frac{m_\chi}{m_\Apr}\right)^4$, with $m_\chi$ being the LDM mass, the following relation can be derived~\cite{Berlin:2018bsc}:
\begin{equation}\label{eq:bound}
    y_{relic} \simeq f \cdot 2\cdot 10^{-14}\left(\frac{m_\chi}{1\,\mathrm{MeV}} \right)^2\; \; ,
\end{equation}
where $f \sim 1$ is a dimensionless quantity that depends on model specific details such as the LDM quantum numbers and the $m_\Apr/m_\chi$ ratio. For a given value of $m_\chi$, it follows from Eq.~\ref{eq:bound} that there is a \textit{target value} of $y$ that experiments should probe, resulting in a clear, predictive target to confirm or rule out the LDM theory~\cite{Battaglieri:2017aum}. 

 Among the experimental techniques adopted to search for vector-mediated LDM at accelerators, the missing-energy strategy 
 has proven particularly effective in the  \textit{invisible decay} scenario ($m_\Apr > 2 m_\chi$)~\cite{Fabbrichesi:2020wbt,Filippi:2020kii,Beacham:2019nyx,Battaglieri:2017aum,Ilten:2022lfq,Graham:2021ggy}. In this approach, the $\Apr$ production signature consists in a large \textit{missing energy}, i.e. the difference between the nominal beam energy and the one deposited in the detector, evaluated event-by-event~\cite{Gninenko:2016kpg}.
For an electron/positron beam setup, two reactions 
are relevant: the radiative and the resonant $\Apr$ production~\cite{Marsicano:2018glj,Marsicano:2018krp}. 
The $\Apr$ radiative production by a high-energy $e^+/e^-$ on a heavy nucleus, $e^\pm Z \rightarrow e^\pm Z \Apr$, scales as $\alpha_{EM}^3\varepsilon^2/m^2_\Apr$, and it is further suppressed at large dark photon mass due to the loss of nuclear coherence and the reduction of the Weiz\"acker-Williams effective photon flux~\cite{Izaguirre:2013uxa}. 
For a narrow $\Apr$ ($\alpha_D \lesssim 0.1$), the cross section $\sigma^P_{res}$ for the annihilation of the incoming 
$e^+$ with an atomic $e^-$, $e^+e^- \rightarrow \Apr$ at peak reads 
\begin{equation}
    \sigma_{res}^P=\frac{1}{\Gamma}\frac{4\pi\alpha_{EM}\varepsilon^2}{m_\Apr} \; \; ,
\end{equation}
where $\Gamma$ is the total $\Apr$ width. As a result, 
the total $\Apr$ yield, obtained by integrating the cross section over the positrons track-length distribution $T(E_+)$~\cite{Chilton}, scales as $N_S \simeq \Gamma \sigma_{res}^{P} T(E_R) \propto T(E_R)/m_\Apr$, where $E_R = m^2_\Apr /(2m_e)$ is the resonant energy. In a recent work, we showed how this mechanism leads to a sensitivity improvement for an \textit{electron-beam} missing-energy measurement, thanks to the secondary positrons of the electromagnetic shower developing in the active target \cite{Andreev:2021fzd}; this effect is maximum for $E_{R}$ close to the missing-energy threshold, and decreases for $E_{R}\rightarrow E_0$ due to the $T(E_+)$ suppression, where $E_0$ is the beam energy. 
In a \textit{positron-beam} missing-energy measurement, instead, thanks to the $T(E_+)$ enhancement for $E_+\rightarrow E_0$ associated to the primary positron, no signal suppression is present, and thus a large sensitivity to $\varepsilon$ is obtained within the $\Apr$ mass range rigidly delimited by the experimental threshold on the missing energy and the energy of the beam~\cite{Marsicano:2018glj}.

\begin{figure*}[t]
\includegraphics[width=.8\textwidth]{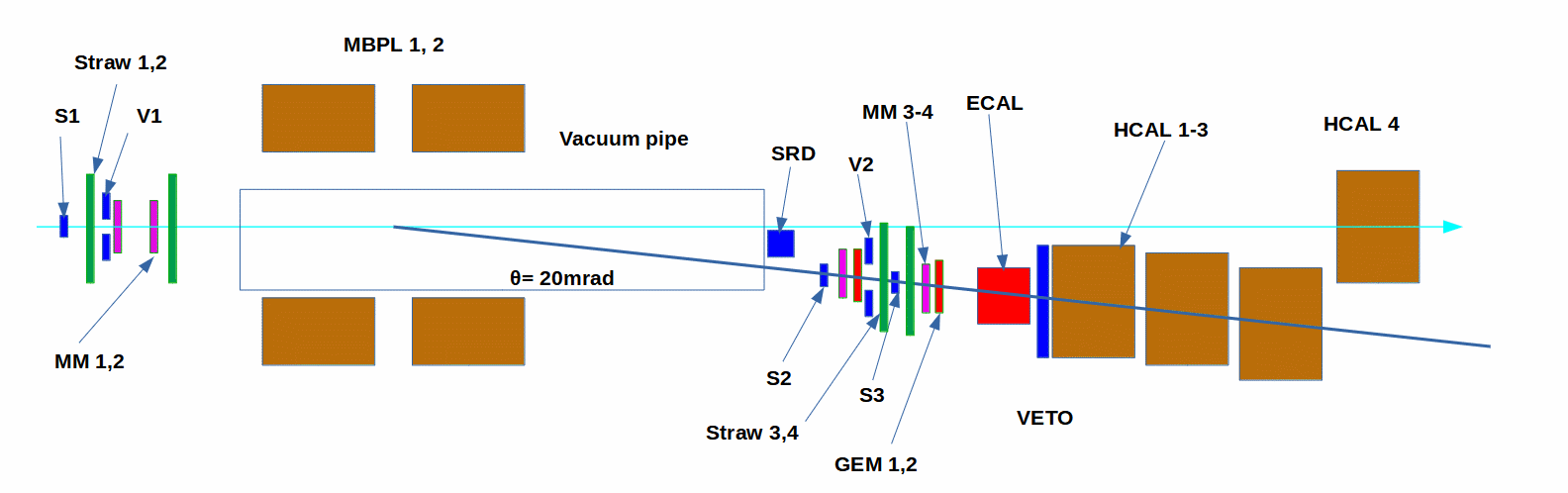}%
{\caption{Schematic illustration of the NA64 setup to search for invisible decays of the $\Apr$s 
resonantly produced by the annihilation of the 100 GeV impinging $e^+$ with the atomic electrons of the active ECAL target.
\label{setup}}}
\end{figure*} 

NA64 is a missing energy experiment exploiting the $100$~GeV electron beam from the H4 beamline~\cite{NA64:2019imj,Andreev:2021fzd}. The H4 beam is produced by the interaction of the primary 400 GeV SPS proton beam with a thick Be target; emerging forward-going photons are pair-converted on a thin lead foil, with a downstream dipole magnet selecting the charge and the momentum of the particles further transported downstream. The beam-line can be also operated in ``positron mode", by appropriately setting the magnets. At 100 GeV/c, the hadronic contamination in positron (electron) mode is $\simeq 4\%$ ($\simeq 0.3\%$)~\cite{Andreev:2023xmj}.
Incoming particles are tagged by a set of three plastic scintillator counters (Sc) and a veto counter. Their momentum is measured by a magnetic spectrometer, composed of a set of tracking detectors (GEMs, MicroMegas, and Straw tubes) installed upstream and downstream a two dipole magnets with total magnetic strength $\int B dl \simeq 7\,$T$\,\cdot\,$m~\cite{Banerjee:2017mdu}. The momentum resolution $\delta p/p$ is $\simeq 1\%$. Particle identification is achieved by measuring the synchrotron radiation (SR) emitted by electrons deflected by the magnetic field through a compact Pb/Sc calorimeter (SRD)~\cite{Depero:2017mrr}. The NA64 active target is a 40$X_0$ Pb/Sc inhomogeneous calorimeter (ECAL), assembled as a $5\times6$ matrix of $3.82\times3.82$ cm$^2$ cells with independent PMT readout, segmented into a $4X_0$ pre-shower section (ECAL0) and a main section (ECAL1).
Downstream the ECAL, a hermetic Fe/Sc hadron calorimeter (HCAL), made by three modules with total length $\simeq 30\lambda_I$, is installed, to detect secondary hadrons and muons produced in the ECAL and upstream detector elements. A high-efficiency plastic scintillator counter (VETO) is installed between the ECAL and the HCAL to further suppress backgrounds. 

In this work, we present the results of the first dedicated positron-beam missing-energy measurement for the search of LDM. This effort paves the way for a future experimental program with positrons, whose sensitivity projection - evaluated in light of this first measurement - is here presented. The analysis is based on a total statistics of about $N_{e^+OT}=(10.1\pm0.1)\times10^{9}$ 100~GeV positrons on target ($e^+OT$)
. A blind-analysis approach was adopted; all selection cuts were optimized by maximizing the signal efficiency for the resonant $\Apr$ production. 
The final cuts configuration required the presence of a well-identified impinging 
track with momentum in the range $[97~\mathrm{GeV},\mathrm{103}~\mathrm{GeV}]$, in time with a total energy deposition in each SRD cell of at least 2.5~MeV. The VETO energy was required to be less than 17 MeV for each panel. A 500 MeV threshold on the ECAL0 energy deposition was applied, and the 
shape of the electromagnetic shower in the ECAL was required to be compatible with that expected for the $\Apr$ signal~\cite{Banerjee:2017hhz}. Finally, the $E_{ECAL}<50~$GeV, $E_{HCAL}<1$~GeV signal region conditions were applied. 

The LDM yield in the signal window was estimated through a full Geant4-based simulation of the NA64 setup, using the DMG4 package~\cite{Bondi:2021nfp}. The signal efficiency was evaluated using data, by applying the analysis cuts on tracking, SRD, VETO, and HCAL to measured 100 GeV $e^+$ events. 
The ECAL shower-shape cut was optimized 
to obtain a 95$\%$ efficiency on measured events where a high-energy $\mu^+\mu^-$ pair is produced in the ECAL mainly from one of the two following production channels: (I) the radiative production, i.e. a Bremsstrahlung photon converting in the EM field of a nucleus, $\gamma N \rightarrow \mu^+\mu^- N$; (II) the atomic annihilation production, $e^+e^-\rightarrow \mu^+\mu^-$.  These ``di-muon'' events, given the MIP nature of high-energy muons, feature a signal-like topology from the point of view of the energy deposition in the~different cells of the ECAL. An additional correction factor, obtained from  Monte Carlo (MC), was introduced for each $m_\Apr$ value to account for residual kinematic differences between signal events and di-muon events. Di-muon events were also used to assess the overall data normalization: the ratio of the observed di-muon absolute yield to the MC-predicted one was $\mathcal{F}_{\mu\mu}=(0.84\pm 0.04)$. In the analysis, we added this as an additional global efficiency factor for the $\Apr$ signal.
Overall, the average signal efficiency was $\xi\simeq (55\pm1)\%$, with a $\simeq 20\%$ dependency on $m_\Apr$ due to the shower-shape cut. The $\xi$ uncertainty was evaluated by varying the cuts used to identify the impinging $e^+$ and assessing the variation of the obtained efficiency. 

\begin{table}[t] 
\begin{center}
\vspace{0.15cm}
\scalebox{.95}{
\begin{tabular}{lr}
\hline
\hline
Background source& Background, $n_b$\\
\hline
(i) $\pi,~K$ decays& $(0.06 \pm 0.03)$ \\
(ii) $e^+$ hadronic interactions in the beam line & $(0.011 \pm 0.007)$\\
(iii) di-muons &  $\leq 0.017$ \\
(iv) $\mu$ decays & $(1.2 \pm 0.2)\times 10^{-3}$ \\
(v) $e^+$ hadronic interactions in the target & $\ll 10^{-3}$ \\
(vi) hadrons interactions in the target & $\ll 10^{-3}$\\
\hline 
Total $n_b$ (conservatively)   & $(0.09 \pm 0.03)$\\ 
\hline
\hline 
\end{tabular}
}
\caption{Expected backgrounds for $N_{e^+OT}=(10.1\pm 0.1)\times10^9$ $e^+OT$.}\label{tab:bckg}
\end{center}
\end{table}

Background events can originate from different sources, summarized in Tab.~\ref{tab:bckg}. The dominant contribution is from the $K^+\rightarrow e^+\pi^0\nu_e$ decay of a kaon contaminant upstream w.r.t. the ECAL,
if the neutrino energy is larger than 50~GeV and the $e^+\gamma\gamma$ particles produce a single low-energy EM shower in the calorimeter. 
The $K^+$ can be misidentified by the SRD cut if a $\delta$-ray is emitted from the $K^+$ through the interaction with upstream beamline materials and then impinges on the SRD detector, or if the $K^+$ is superimposed in time with a low-energy $e^+$ from the beam tail, that emits enough SR and is then deflected away by the magnets. A lesser contribution comes from the branching-ratio suppressed $\pi^+\rightarrow e^+\nu_e$ decay.
These backgrounds, evaluated from data, are potentially critical for a positron measurement at the H4 line, since the hadronic contamination is significantly larger when compared to the electron beam case~\cite{Andreev:2023xmj}.
The second most intense background channel is the upstream hadrons production by the $e^+$ interaction with beamline materials~\cite{Gninenko:2016kpg}, when the low-energy positron is measured by the ECAL, and one or more high-energy neutral hadrons escape detection. 
We estimated this contribution from a larger dataset collected by NA64 in 2022 using a 100 GeV electron beam, with the same detector configuration and reversed beamline optics, following the method described in Ref.~\cite{Banerjee:2017hhz}. 
The remaining sub-dominant background sources include: (III) the loss of di-muon events, evaluated mainly from data by accounting for the VETO inefficiency ($\simeq 1.4\%$) and for the probability of the di-muons deposited HCAL energy to fall below 1 GeV ($\simeq 5\times10^{-6}$); (IV) the in-flight decay of a  muon contaminant, evaluated from Monte Carlo simulations; and (V - VI) the probability for a positron or a hadron contaminant to undergo a hard interaction in the ECAL, depositing therein less than 50 GeV and producing final-state particles all undetected by the VETO and the HCAL. Summing up all these contributions, a total background yield of $B=(0.09\pm0.03)$ events is expected.

\begin{figure}
    \centering
    \includegraphics[width=.48\textwidth]{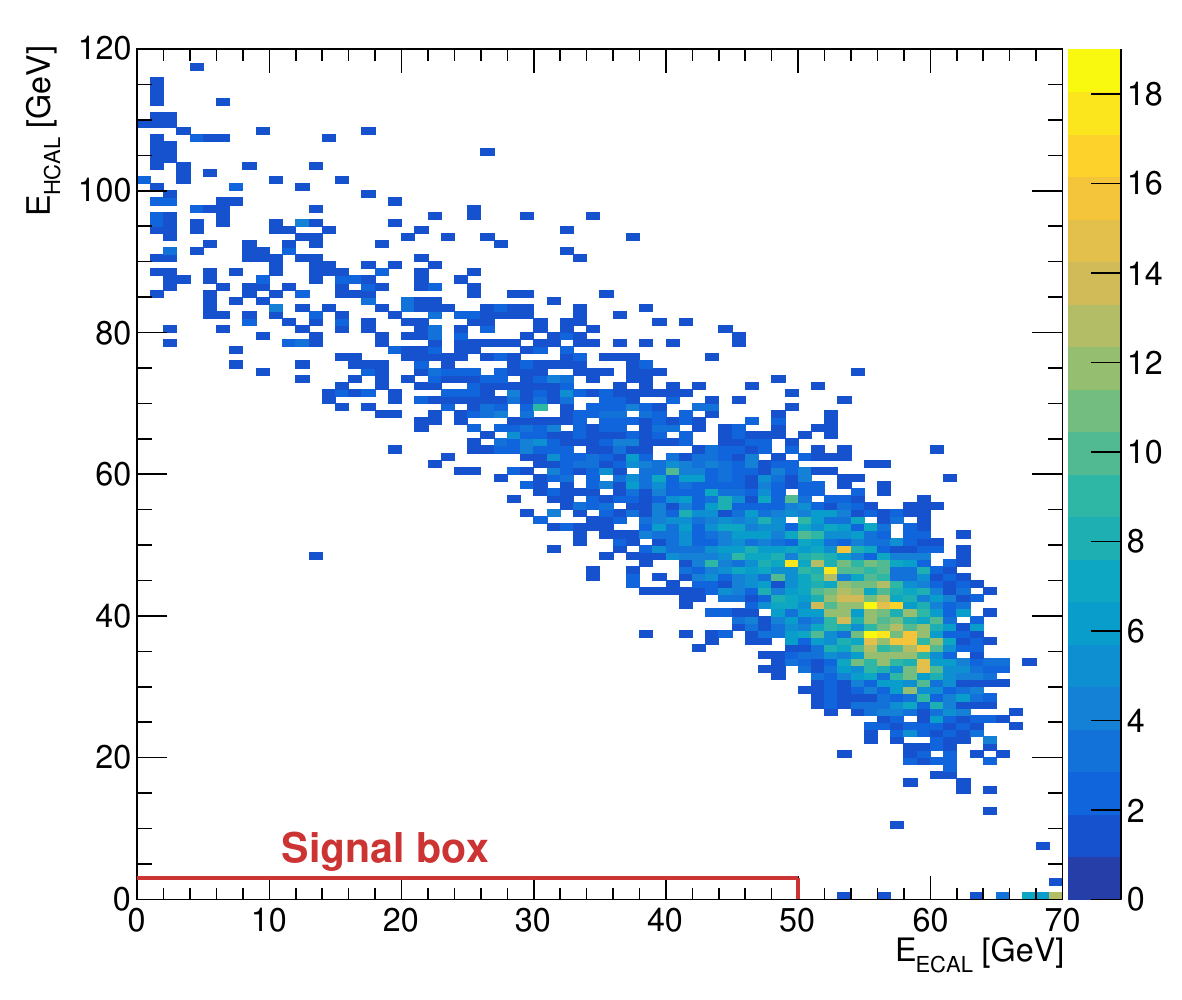}
    \caption{The unblinded ECAL vs HCAL energy distribution for events satisfying all analysis cuts. The unblinded signal region is also reported -- for better visualization, it has been expanded by a factor $\times 3$ along the vertical axis.}
    \label{fig:data}
\end{figure}

The ECAL vs HCAL energy distribution for the events satisfying all analysis cuts is reported in Fig.~\ref{fig:data}. The diagonal band corresponds to events from the reaction $e^+ N\rightarrow e^+ X$ in the ECAL, with final state hadrons interacting with the HCAL. After unblinding, no events have been observed in the signal region. Based on this result, we derived the upper limit on the $\Apr$ coupling $\varepsilon$ as a function of $m_\Apr$. We adopted a frequentist approach, considering the $90\%$ Confidence Level (CL) of a one-sided profile-likelihood test statistics~\cite{Gross:2007zz}. The likelihood model was built assuming a Poisson PDF for the number of events in the signal region, with mean value $\mu=S+B=(\varepsilon/\varepsilon_0)^2S_0+B$, where $S_0$ is the signal yield for the nominal coupling value $\varepsilon_0$ obtained from Monte Carlo.

\begin{figure}[t]
    \centering
    \includegraphics[width=.49\textwidth]{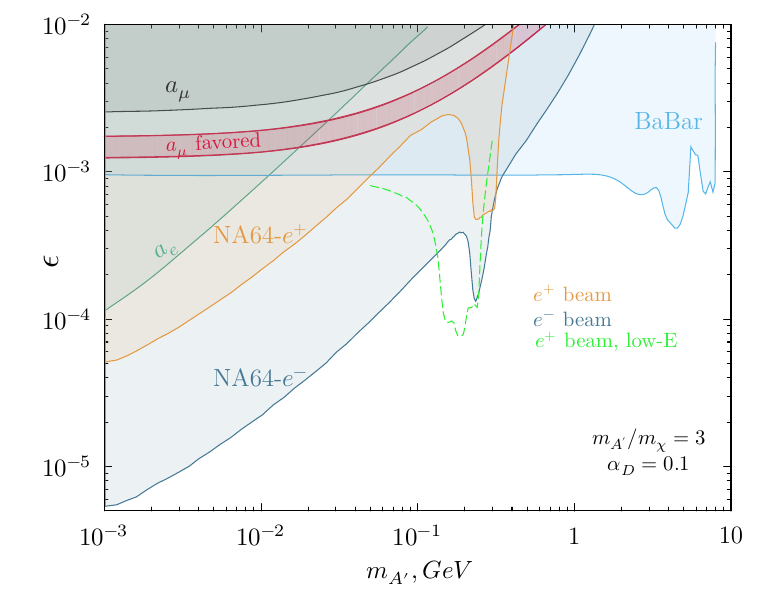}
    \caption{    
    The new exclusion limits from the positron-beam, missing energy measurement presented in this work in the $\varepsilon$ vs $m_\Apr$ space, considering fermionic LDM for $\alpha_D=0.1$. The most stringent LDM exclusion limits from BaBar~\cite{Lees:2017lec} and NA64~\cite{Andreev:2023uwc} are also shown, as well as the favored area from the muon $g-2$ anomaly~\cite{Pospelov:2008zw,Abi:2021gix} (red lines). The green dashed lines reports the sensitivity for a future positron-beam effort at lower energy, as described in the text. For a full review on other exclusion limits and planned measurements see Ref.~\cite{Fabbrichesi:2020wbt,Filippi:2020kii,Beacham:2019nyx,Battaglieri:2017aum,Ilten:2022lfq,Graham:2021ggy}. }
    \label{fig:limit_eps}
\end{figure}

The systematic uncertainties were accounted for in the statistical procedure by introducing, for each contribution, an additional log-normal PDF term in the Likelihood, taking the measured value as the \textit{observed} one, and handling the \textit{expected} value as a nuisance parameter~\cite{Gross:2007zz}. The uncertainty on the number of expected background events (30$\%$) and that on the signal yield from the number of $e^+OT$ ($0.1\%$), from the tracking, SRD, VETO, and HCAL cuts ($\simeq 1\%$), and from the overall normalization as obtained from the di-muons analysis ($4.8\%$) are the same for all $m_\Apr$ values. In contrast, the systematic uncertainty arising from the ECAL energy calibration, and thus affecting the pre-shower and missing energy thresholds, result in a $m_\Apr$-dependent effect. For the missing-energy threshold, we estimated $\delta_E/E \simeq 1.5\%$ by measuring the position of the full-energy peak in the calorimeter across the different runs. 
For the pre-shower threshold, instead, we evaluated $\delta_E/E \simeq 1.0\%$ from the analysis of di-muon events. To evaluate the impact of $\delta_E/E$ on $S_0$, first we computed the signal yield for the nominal threshold values. Then, we randomly sampled these multiple times from two Gaussian distributions with $\mu=50$~GeV and $\sigma=0.75$~GeV and $\sigma=0.5$~GeV, respectively, computing $S_0$ for each configuration and then taking the RMS of all results as the systematic uncertainty.
The largest systematic uncertainty, $\simeq 25\%$, is observed for the dark photon resonant production at $m_\Apr\simeq225$ MeV, since in this case the resonant energy is close to the missing energy threshold.

\begin{figure*}[t]
    \centering
    \includegraphics[width=.49\textwidth]{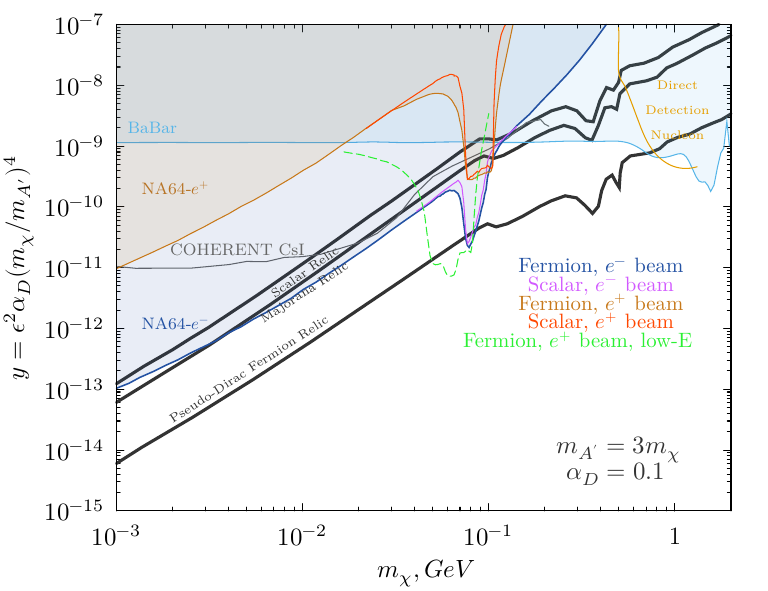}
    \includegraphics[width=.49\textwidth]{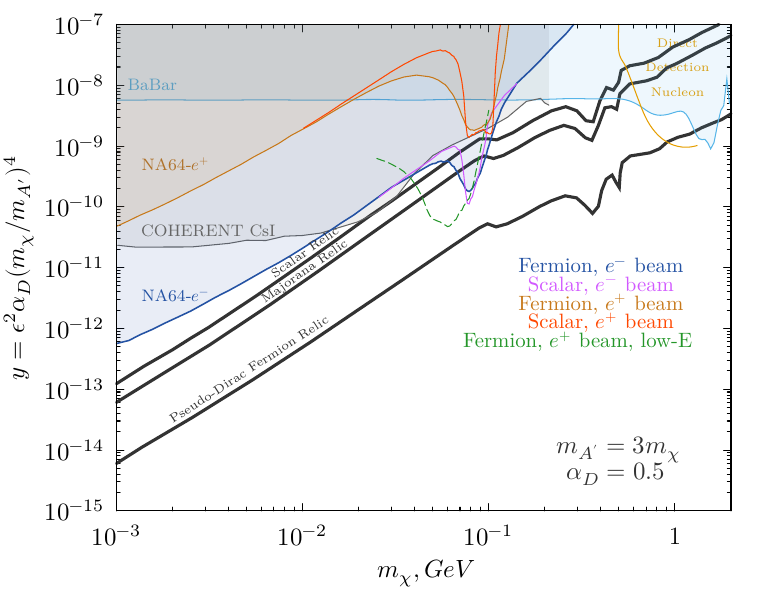}
    \caption{The new NA64 positron-beam exclusion limit in the $(m_\chi,y)$ plane, for $\alpha_D=0.1$ (left) and $\alpha_D=0.5$ (right). The other curves and shaded areas report already-existing limits in the same parameter space from NA64 in electron-beam mode~\cite{Andreev:2023uwc}, COHERENT~\cite{COHERENT:2021pvd}, and BaBar~\cite{Lees:2017lec}.
    The black lines show the favoured parameter combinations for the observed DM relic density for different model variations~\cite{Berlin:2018bsc}. In each panel, the green dashed line shows the sensitivity for a future positron-beam effort at lower energy.}
    \label{fig:limit_y}
\end{figure*}

The obtained upper limit on $\varepsilon$ as a function of $m_\Apr$ is shown in Fig.~\ref{fig:limit_eps} for fermionic LDM, with $\alpha_D=0.1$ and $m_\Apr = 3 m_\chi$.
In the same figure, the combined projection for a future measurement program at two different beam energies - 60 GeV and 40 GeV - is presented. For both measurements, a statistics of $10^{11}\,\,e^+$OT was considered, assuming a missing energy threshold corresponding to half the energy of the beam and a $50\%$ signal efficiency, in analogy with the strategy adopted for the 100 GeV data analysis. Limits were evaluated considering no expected background events; this assumption is corroborated by the fact that the main background source for such measurements comes from the hadronic contaminant fraction in the beam, whose intensity wanes 
at decreasing beam energy~\cite{Andreev:2023xmj}. The same results are reported in Fig.~\ref{fig:limit_y} as a function of $y$. The black lines correspond to the aforementioned ``targets'' predicted by cosmology, in the hypothesis of a thermal LDM origin. To assess the variation of our result as a function of the LDM model parameters, we repeated the upper limit calculation considering also a scalar LDM model and the two $\alpha_D$ values, 0.1 and 0.5.

In conclusion, we performed the first positron-beam missing-energy measurement searching for LDM. The resulting limits touch our latest electron-beam results, corresponding to $\sim$2 orders of magnitude larger electron statistics~\cite{Andreev:2023xmj}. This proves the outstanding potential of NA64 to probe dark sectors using the unique H4 positron beam at the CERN SPS, and motivates a multi-energy measurement program to fully exploit the peculiarities of the resonant $\Apr$ production to ``scan'' the LDM parameter space. Such experimental program would allow to explore a significant region of the LDM parameter space, in synergy with our well established electron-beam measurements.

\begin{acknowledgments}
We gratefully acknowledge the support of the CERN management and staff, and the technical staffs of the participating institutions for their vital contributions. 
This result is part of a project that has received funding from the European Research Council (ERC) under the European Union's Horizon 2020 research and innovation programme, Grant agreement No. 947715 (POKER). 
This work was supported by the HISKP, University of Bonn (Germany), ETH Zurich and SNSF Grant No. 169133, 186181, 186158, 197346 (Switzerland), and ANID - Millenium Science Initiative Program - ICN2019 044 (Chile), and  RyC-030551-I and PID2021-123955NA-100 funded by MCIN/AEI/ 10.13039/501100011033/FEDER, UE (Spain).
\end{acknowledgments}


\bibliographystyle{apsrev4-2}

\bibliography{../Bibliography/bibliographyNA64_inspiresFormat.bib,../Bibliography/bibliographyNA64exp_inspiresFormat.bib,../Bibliography/bibliographyOther_inspiresFormat.bib} 

\end{document}